\documentstyle[aps,eqsecnum,epsf,epsfig]{revtex}

\begin{document}

\baselineskip=16pt
%\draft

\title{Scenario for Ultrarelativistic Nuclear Collisions: \\
  II.~  Geometry of quantum states at the earliest stage. } 
\author{ A.  Makhlin }
\address{Department of Physics and Astronomy, Wayne State University, 
Detroit, MI 48202}
\date{July 26, 2000}
\maketitle
\begin{abstract}

We suggest  that the ultrarelativistic collisions of heavy ions provide the
simplest situation for the study of strong interactions which can be understood
from first principles and without any model assumptions about the microscopic
structure of the colliding nuclei. We argue that the boost-invariant geometry
of the collision, and the existence  of hard partons in the final states, both
supported by the data, make a sufficient basis for the quantum theory of the
phenomenon. We conclude that the quantum nature of the entire process is
defined by its global geometry, which is enforced by a macroscopic finite size
of the colliding objects.  In this paper, we study the qualitative aspects of
the theory and  review its development in two subsequent papers. Our key
result is that the effective mass of the quark in the expanding system formed
in the collision of the two nuclei is gradually built up reaching its maximum by
the time the quark mode becomes sufficiently localized. The 
chromo-magneto-static interaction of the  color  currents flowing in the
rapidity direction is the main mechanism which is responsible for the 
generation of the effective mass of the soft quark mode and therefore, for the
physical scale at the earliest stage of the collision. 

\end{abstract}

\section{Introduction}
\label{sec:S1}

In our previous papers \cite{QFK,QGD,tev}, we began a systematic theoretical
study of the scenario of ultrarelativistic collision of heavy ions.   Our main
result obtained in Ref.~\cite{tev} (further quoted as paper [I]) was that the
dense system of quark and gluons which is commonly associated with the
quark-gluon plasma (QGP) can be formed only {\em in a single quantum
transition}. In this and two subsequent papers we continue to develop this
approach in greater detail. We come to a conclusion that ultrarelativistic
nuclear collisions is a unique physical phenomenon when the quantum dynamics of
the process is enforced by a macroscopic finite size of the colliding objects
rather than by a microscopic origin of their constituents. 

The entropy (the number of excited degrees of freedom) produced in collisions
of heavy ions is a natural measure of the strength of the  colored fields
interaction. Indeed, before the collision, the quark and gluon fields are
assembled into  two coherent wave packets (the nuclei) and therefore, the
initial entropy equals zero. The coherence is lost, and  entropy is created due
to the interaction. A search for the QGP in heavy-ion collisions is, in the
first place, a search for evidence of  entropy production.  Though one may wish
to rely on the invariant formula  $S={\rm Tr}\rho\ln\rho$,  which expresses the
entropy $S$ via the density  matrix $\rho$, at least one basis of states should
be found explicitly.  It is imperative to design  such a basis, and to
practically study the collective effects that take place at the earliest
($<1fm$) stage of the collision.

 In any standard {\em exclusive} scattering process, no entropy can be produced
since the scattering process begins with  a pure quantum state of two stable
colliding particles and the final state is also given as a pure state of
several particles in exactly known quantum states. The only way one can address
the quantum problem of entropy production is to consider {\em inclusive}
measurements. Since these measurements are not complete (i.e., are not
exclusive), they indeed form an ensemble with finite entropy.

Quantum chromodynamics still cannot provide the theory of nuclear collisions
with detailed information about nuclear structure before the collision. We
face a formidable task  to build a reliable theory of nuclear collision knowing
almost nothing about the initial state. We may rely safely  only upon the fact
that the nuclei are stable bound states of the QCD and therefore, their
configuration is dominated by the stationary quark and gluon fields which are
genuine constituents of these quantum states.  Fortunately, this,  at first
glance  very scarce, information appears to be sufficient for the understanding
of many intimate details of the collision process, if the problem is addressed
from  first principles.

We hope that  the final state is defined more accurately and {\em believe} that
a single-particle distribution of quarks and gluons at some early moment after
the nuclei have intersected, describe it sufficiently. Thus, we may count upon
a reasonably well-defined quantum observable. The measurement of the
one-particle distribution is an inclusive measurement.  The corresponding
operator should count the number of final-state particles defined as the
excitations above the perturbative vacuum.   As long as we expect that this
counting makes sense  on the event-by-event basis, the collision is indeed
producing the entropy. To develop the theory for this transition process we
have to cope with a binding feature that the ``final'' state has to be defined
at a finite time. This may look disturbing for readers well versed in
scattering theory, because the whole idea of {\em a scenario as a temporal
sequence of different stages} is alien to the standard S-matrix theory. The
general framework of an appropriate theory, named quantum field kinetics (QFK),
has been developed in  our previous papers\cite{QFK,QGD,tev}. It is based on a
remarkable similarity; the measurement of one-particle distributions is as
inclusive as the measurement of the distribution of the final-state electron in
deeply inelastic {\em ep}-scattering (DIS). This  conceptual similarity,
however, meets difficulties in its practical implementation. 

~ (i)  The inclusive DIS directly measures only the  electromagnetic 
fluctuations in the proton.  The problem is posed according to the  S-matrix 
scattering  theory improved by means of the renormalization group. The concept
of   running coupling emerges  precisely in this context. The  operator product
expansion  (OPE) allows one to hide all the unknown information  about the
proton (commonly associated with large distances) into the local  operators of
various dimensions.  Introduced in this way, structure functions  (given
explicitly in terms of their momenta) are applicable to DIS process, and  only
to DIS.

~(ii)  It is impossible to derive structure functions of {\em pp}-scattering
(not to say about  {\em AA} ) using the  OPE method, because in this  case the
composite QCD operators become essentially non-local.\footnote{ Similar 
situation takes place in the {\em ep}-process if a jet is chosen  as the
inclusive  observable. Then the dynamics of the process is sensitive to the QCD
content of the electron (see Sec.IV in paper [I]).}  

~(iii)  Historically, the escape was provided by the parton model ( the 
factorization hypothesis ), which was successfully applied to various
processes   that accompany {\em pp}-scattering (like Drell-Yan pairs
production), where the  factorization scale can be  kept under the data
control, since the number of  particles in the final state is relatively
small.  In {\em AA}-collisions, the  control over the factorization scale is
practically impossible because of the enormously  high multiplicity.
Furthermore, the phase space of the final state is densely  populated and the
picture of an independent emission (unitary cut in the Feynman diagram)
employed for the  derivation of DIS structure functions does not hold any more.

 As it has been already mentioned, in the {\em AA}-case we need a 
developing in time scenario which cannot be accessed from the S-matrix 
scattering theory, while the DIS structure functions are constructed within 
S-matrix approach. The QFK method has been developed in order to resolve these
problems by  addressing not only the nuclear collision as a transient process,
but the  {\em ep}-DIS also. Our major hope was to derive QCD evolution
equations and to  introduce the structure functions using the framework of an
independently initiated theory of nuclear collisions. The first step along 
this guideline was an immediate success \cite{QFK,QGD}. It was demonstrated, 
that the evolution equations indeed describe a transient process that ends as 
an electromagnetic fluctuations inclusively probed by the electron.

To give a flavor of how the method works practically, let us start with a
qualitative description of the inclusive {\em e-p} DIS measurement (for now,
at  the tree level without discussion of the effects of interference). In this
experiment, the only observable is the number of electrons with a given
momentum in the final state. Something {\em in the past} has to create the
electromagnetic field that deflects the electron. {\em Before} this field is
created, the electromagnetic current, which is the source of this field, has to
be formed. Since the momentum transfer in the process is very high, the current
has to be sufficiently localized. This localization requires, in its turn, that
the electric charges which carry this current must be dynamically decoupled
from the bulk of the proton {\em before} the scattering field is created (to
prevent a recoil to the other parts of the proton which could spread the
emission domain). Such a dynamical decoupling of a quark requires a proper
rearrangement of the gluonic component of the proton with the creation of
short-wave components of a gluon field. By causality, corresponding gluonic
fluctuation must happen {\em before} the current has decoupled, etc.  Thus we
arrive at the picture of the sequential-dynamical fluctuations which create an
electromagnetic field probed by the electron.  The lifetimes of these
fluctuations can be very short. Nevertheless, they all {\em coherently} add up
to form a stable proton, unless the interaction of measurement breaks the
proper balance of phases. This intervention freezes some instantaneous  picture
of the fluctuations, but with wrong ``initial velocities'' which  results in a
new wave function, and collapse of the old one. This  qualitative picture has
been described many times and with many variations  in the literature, starting
with the pioneering lecture by Gribov  \cite{Gribov}, and including a recent
textbook \cite{BPQCD}; however, the sequential temporal ordering of the
fluctuations has never been a key issue. We derive this ordering as a
consequence of the Heisenberg equations of motion for the observables. The
practical scheme of calculation that emerges in this way appears to be {\em a
special form of quantum mechanics which describes an inclusive measurement as a
transient process}. Translated into mathematical language in momentum space,
this picture leads to the most general form of the evolution equations, which
may then be reduced (under different assumptions) to the known DGLAP, BFKL, or
GRV equations \cite{QGD}. The evolution equations were derived  immediately in
the closed form of the integral equations avoiding a selective  summation of
the perturbation series. The standard inclusive {\em e-p} DIS  indeed delivers
information about quantum fluctuations which may {\em  dynamically} develop in
the proton {\em before} it is destroyed by a hard  electromagnetic probe. One
of the most amazing features that has been discovered  in the framework of QFK
is that the QCD evolution equations are an {\em intrinsic  property of the
inclusive measurement process}, and they are not limited by the  factorization
condition.

In paper [I], we studied the problem of loop corrections in the QFK  evolution
equations. First, we found that they do not corrupt the  causal picture of the
measurement described above, at the tree level. Second, they indeed provide a
{\em scale} to the  entire process. This scale is connected with collective
interactions in the final state, which  dynamically generate masses for the
final states of emission, thus  regulating the abundant collinear divergences
of the null-plane dynamics. We  required that the real parts of all radiative
corrections (phase shifts)  must vanish along the direction of the
initial-state propagation of the  colliding objects. Thus, we explicitly
accounted for the integrity of the  nuclear wave function {\em before} the
collision. This special choice of the  renormalization point, is {\em natural}
for ultrarelativistic nuclear  collisions, since it allows one to treat
nuclei as finite-size quantum  objects and incorporate their Lorentz
contraction as a classical boundary  condition imposed on the space-time
evolution of quantum fields after the  nuclear coherence is broken. 

The net yield of our previous study in paper [I] can be summarized as
follows: The interaction between the two ultrarelativistic nuclei switches
on almost instantaneously. This interaction explores all possible quantum
fluctuations which could have developed by the moment of the collision and
freezes (as the final states) only the fluctuations compatible with the
measured observable. These snapshots cannot have an arbitrary structure,
since the emerging configurations must be consistent with all the
interactions which are effective on the time-scale of the emission process. 
In other words, the modes of the radiation field which are excited in the
course of the nuclear collision should be the collective excitations of the
dense quark-gluon system. This conclusion is the result of an intensive
search of the {\em scale} inherent in the process of a heavy-ion collision.
We proved that the scale is determined only by the physical properties of
the final state. 

Our previous study clearly indicates that a theory that describes both
phenomena (i.e. ep-DIS and AA-collisions) from a  common point of view can be
built on two premises: causality, and the  condition of emission.  The latter
is also known as the principle of cluster  decomposition, which must hold in
any reasonable field theory. What the  ``resolved clusters'' are is a very
delicate question. These states should  be defined with an explicit reference
as to how they are detected.  Conventional detectors deal with hadrons and
allow one to hypothesize about  jets. QGP turns out to be a kind of collective
detector for quarks and  gluons. In DIS experiment, the new wave function is
measured {\em inclusively}  which itself could be the source of the entropy
production if the final  state had some properties of the collective system.
This collective  system would then be a detector. It turns out that a dramatic
difference  in population of the final states is the sole fact that makes DIS
and heavy  ion collisions so different.

At this moment, the natural line of development of these ideas brought 
us to the point when any further progress is impossible without 
explicit knowledge of the {\em normal modes of the expanding dense 
quark-gluon system}. There are several conceptual and technical 
problems of different caliber where this knowledge is crucial. 

~{\em 1.}  Most of the entropy is expected to be produced during the  initial
breakup of the nuclei coherence. Computing the entropy amounts  to the digital
counting of the exited degrees of freedom. Therefore, the  states themselves
must be precisely defined. From this point of view, the role of dynamical
masses of the normal  modes is decisive. They provide  an infrared boundary
for the space  of final states thus making the possible number of the excited 
states (the entropy) finite. 

~{\em 2.}  Only after the infrared boundary for the QCD states is  found can
we hope to have a self-consistent perturbation theory.  This was an original
idea which motivated the search of the QGP  \cite{Shuryak}. A perturbative
description at the kinetic stage of  the scenario cannot rely on massless  QCD,
which has no intrinsic scale. It can be effective only if it is
based on the interaction  of the partons-plasmons, i.e., quarks and gluons
with the effective  masses. Built on these premises,  the scenario for the
ultra- relativistic nuclear collision promises to be more perturbative  than
the standard pQCD. 

~{\em 2.} Standard perturbative calculations with massless gauge fields 
always lead to collinear singularities that require a parameter of  resolution
for their practical removal. When this singularity is due  to the emission
into the final state, then this parameter is usually found as a property of
the detector. As long as we consider the QGP itself as a detector, no external
parameters of this kind can be in the theory. Collinear problems also appear
in  loop corrections, even in their imaginary parts. Therefore, they are also
due to real processes. [ In physical gauges, the collinear singularities in the
loop corrections can also be connected with the spurious poles of the gluon
propagator, which is a consequence of an incomplete gauge fixing and imperfect
separation between the longitudinal fields and the fields of radiation.]

As it has been discussed in paper [I], the collinear problems in perturbative
QCD  show up only because the unphysical states are added  to the list of the
possible final states of the radiation processes. These states can be
eliminated from the theory by accounting for the real interactions in the
final state which provide effective masses for all radiated fields. In the 
null-plane dynamics, this appeared to be impossible, since any type of kinetics
that may lead to the formation of the effective mass is frozen on the light
cone. In order to have  meaningful evolution equations for heavy ion collisions
we must account for the dynamical masses of the realistic final states in dense
expanding matter; the QCD evolution has to provide a kind of self-screening of
the collinear singularities. The way the effective mass was {\em estimated} in
paper [I] was crude, and it was our original goal to improve the calculations
using  the full framework of {\em  wedge dynamics}.

\section{Outline of ideas, calculations, main results, and conclusion}
\label{sec:SNo}

In this section, we  review the work presented in this and two subsequent
papers \cite{geg,fse} (hereafter quoted as papers [III] and [IV]). Our approach
is strongly motivated by an idea, that collisions of ultrarelativistic heavy
ions is the cleanest laboratory where one can study the dynamics of strong
interactions. We consider an adequate choice of the interacting quantum states 
at different stages of the scenario as the issue of first priority. The focus
of our previous study was on the QCD evolution equations in the environment of
the heavy ion collision. Now, we concentrate on the possible properties of the
state that emerges immediately {\em after} the coherence of the nuclei is
broken by the first hard interaction. As in paper [I], we view the dynamics of
the early stage as a single quantum process and concentrate on the study of
quantum fluctuations subjected to the condition of a simple inclusive
measurement (currently, on the inclusive one-particle distributions). We
endeavor to take full advantage of approaching the problem from  first
principles.

\subsection{Heuristic arguments.}
\label{subsec:So0} 

Collision of ultrarelativistic heavy ions is such a unique physical phenomenon,
that it is difficult to find its complete analog throughout everything that
has been studied in physics previously. However, we can point to several
examples which share some common distinctive patterns  with the process under
investigation. We begin with these examples in order to help the reader 
understand the ideas of our new synthesis. 

{\em 1}. Let an electron-positron pair be created by two photons. If the energy
of the collision is large, then the electron and positron are created in the
states of freely propagating particles and the cross-section of this process
accurately agrees with the tree-level perturbative calculation. However, if the
energy of the collision is near the threshold of the process, then the relative
velocity of the electron and positron is small, and they are likely to form
positronium.  It would be incredibly difficult to compute this case using
scattering theory.  Indeed, one has to account for the multiple emission of
soft photons which gradually builds up the Coulomb field between the electron
and positron and binds them into the positronium. However, the problem is
easily solved if we realize that the bound state {\em is} the final state for
the process. We can still  use low-order perturbation theory to study the
transition between the two photons and the bound state of a pair
\cite{Sakharov}.

{\em 2}. Let an excited atom be in a cavity with ideally conducting walls. The
system is characterized by three parameters: the size $L$ of the cavity, the
wave-length $\lambda\ll L$ of the emission, and the life-time $\Delta
t=1/\Gamma$ of the excited state. The questions are, in what case  will the
emitted photon bounce between the cavity walls, and when will the emission
field be one of the normal cavity modes? The answer is very simple. If $c\Delta
t\ll L$, the photon will behave like a bouncing ball.  When the line of
emission is very narrow, $c\Delta t\gg L$, the cavity mode will be excited. It
is perfectly clear that in the first case, the transition current that emits
the photon is localized in the atom. In the second case it is not.  By the time
of emission, the currents in the conducting walls have to rearrange charges in
such a way that the emission field immediately satisfies the proper boundary
conditions. We thus have a collective transition in an extended system.

{}From a practical point of view, these two different problems are united by
the method of obtaining their solutions. A part of the interaction (Coulomb
interaction in the first case, and the interaction of radiation with the cavity
walls in the second case) is attributed to the new ``bare'' Hamiltonian which
is diagonalized by the wave functions of the final state modes. The less
significant interactions can be accounted for by means of perturbation
theory. For us, the most important message is that it is possible to avoid a
difficult study of the transient process that physically creates these modes.

{\em 3}. Let an experimental device consists of quantum detectors that 
register photons emitted by a pulse source. Each pulse initiates an
``event''.   Let a sheet of glass is placed somewhere between the source and
detectors. If this glass were installed permanently in a fixed position, then
the method to account for its presence would be trivial. One must expand the
field of the initial light pulse over the system of modes (Fresnel triplets of
incident, reflected, and refracted waves) that satisfy the continuity
conditions on the glass boundaries. The quantum theory would then treat these
triplets as the photons, etc. When the position of the glass sheet is unknown,
e.g., it changes in the time periods between the pulses, then such a universal
decomposition becomes impossible. Nevertheless, in each particular event there
exists an important element of {\em classical boundary conditions}. Using
special tricks (e.g., by measurement of the times of arrival of the
precursors), one may determine the glass position and thus to learn how the
translational symmetry of free space was actually broken and what are the
photons of a particular event. Though the whole set up of this example is
artificial, it illustrates the major idea.  The quantum theory of an individual
event can be fully recovered, even if macroscopic parameters of the theory
are not known until the event is completely recorded.  Indeed, the prepared at
a large distance light pulse can be expanded over any of the systems of the
Fresnel triplets (corresponding to different positions of the glass sheet).
Only after analysis of the data can it be learned, which of these
decompositions is meaningful. One can fill the space between the detectors with
gas and account for the interaction between the light and gas (or even for a
non-linear interaction of photons in the gaseous medium) by perturbation
theory. Being the non-interacting waves in ``free space'', the Fresnel triplets
will serve as the zeroth-order approximation of a quantum theory.  One may also
decide not to begin with the triple waves. Then the glass must be treated as 
an active element. The same triplets will be recovered in the course of a real
transient process on the glass surface. The  translational symmetry will be
broken dynamically.

A very similar picture develops during the heavy ion collision. The normal
modes of the final state are formed in the course of real interactions. The
mechanism responsible for effective mass of the plasmons is illustrated by the
first two examples. The third example points us to an optimal choice of the
zeroth-order approximation. Exactly in the same way as the reflected and
refracted waves of the Fresnel triplet cannot physically appear before the
light front reaches the glass surface, nothing can happen with the nuclei
before they overlap geometrically. Only at this instance the interaction
determines the collision coordinates in $(tz)$-plane. The symmetry gets
uniquely broken, and the normal modes of the propagating colored fields after
the interaction exist only inside the future region of the interaction domain.
If the coupling is small then we may disregard  later interactions. However,
the system of the final-state free fields will have a broken translational
symmetry, which will be remembered by the normal modes that obey certain
macroscopic boundary conditions.

Referring to the above examples, one should keep in mind the source of the
major difference between the QED and QCD phenomena. The local gauge symmetry of
QED  can be extended to a global gauge symmetry which generates the conserved
global quantum number (electrical charge) which can be sensed at a distance.
The proper field of an electric charge is the main obstacle for the definition
of its size. On the other hand, the radiation field of QED appears as a result
of  the changes in the extended proper fields of accelerated charges, and one
can physically create such an object as a front of electromagnetic wave. In QCD,
the local gauge invariance of the color group does not correspond to any
conserved charge. Hence, we can easily determine the size of the colorless
nucleus, but we cannot create a front of color radiation in the gauge-invariant
vacuum. These two properties of QCD both work for us. They allow one to use
the Lorenz contraction to localize the initial moment of the collision and
thus, to impose the classical boundary conditions on the propagating color
fields at the later times. The existence of the collective propagating quark
and gluon modes at these times is the conjecture that has to be verified by the
study of heavy ion collisions.

The finite size of the colliding nuclei, and a strong localization of the
initial interaction as its consequence, is a sufficient input for the theory
that describes the earliest stage of the collision. The formalism of quantum
field theory appears to be a powerful tool that allows one to derive many
properties of the quark-gluon system after the collision.

\subsection{Phenomenological input.}
\label{subsec:So1} 

~~(i)~ We consider the rapidity plateau seen event-by-event in nuclear 
collisions at very large energy as a confirmed by the data indication that the
quantum transient process has no scale corresponding to the finite resolution
in the $t$- and $z$-directions. By a common wisdom, the absence of this scale
must cause the boost-invariant expansion.\footnote{This plateau in the
distribution of the final-state hadrons  is clearly seen even in the inclusive
jet distribution in $ep$-DIS data, but only statistically.} 

~~(ii)~ All existing data indicate that, regardless of the nature of the
colliding objects, a certain number of particles with large transverse momentum
are found in the final state. At high $p_t$, the cross section reasonably well
follows the Rutherford formula. We rely on the universality of  Rutherford 
scattering as an indication that there is no scale parameter of resolution in
the transverse $(xy)$-plane that characterize this process. We assume that in
nuclear collisions, these hard states created at  the very early instance of
the collision can be described by the one-particle distribution measured on
event-by-event basis.

\subsection{Ideas.}
\label{subsec:So2} 

~~(i)~ The finite size of colliding nuclei plays a crucial role in our approach
since it allows for a realistic measurement of  the Lorentz contraction thus
precisely fixing the time and the coordinate of the collision point. In the
laboratory frame, both nuclei are Lorentz contracted to a longitudinal size
$R_0/\gamma \sim 0.1 fm$, while the scale relevant for the hadron structure is 
$ \sim 0.3 fm$. Therefore,  in the center-of-mass frame, both nuclei are
passing through a ``pin-hole'', and the detailed information about the
microscopic nuclear structure is not essential. A precise measurement of the
{\em velocity}, i.e. the coordinates at two close time moments, is impossible
\cite{Landau}. Hence, a celebrated rapidity plateau in every single collision
of two ultrarelativistic ions is a direct consequence of this type of
measurement. We accept the fact of the rapidity plateau as a classical boundary
condition for the quantum sector of the theory.

~~(ii)~ There is no doubt that the entire collision process must develop
inside the future light cone of the collision domain. Only there can the
dynamics of the propagating color fields become a physical reality. In other
words, the resolution of colored  degrees of freedom is a consequence of the
precise measurement of the coordinate by means of strong interactions. 

~~(iii)~ The true scale of the entire quantum process coincides with its
infrared boundary, which is build dynamically in the course of this process.
Namely, the hard partons, which are produced as localized and countable
particles at the earliest time of the process, define masses for the soft field
states formed at the later times thus bringing the transient process to its
saturation.

\subsection{Strategy and theoretical foundations.}
\label{subsec:So3} 

Addressing the problem of normal modes of the expanding quark-gluon system,
we 
proceed in two major steps. First, we study the classical and quantum
properties of the normal modes subjected to the boundary condition of a
localized interaction that follows from the relativistic causality (being
the free fields in all other respects). Then, we use these modes as a basis
for the perturbation theory and compute the effective mass of the quark
propagating through the background distribution of hard partons.

~~(i)~ We begin in Sec.~\ref{subsec:SBe1} with the qualitative study of free
fields, fully incorporating the properties of the geometric background of the
expanding matter. Taking the simplest plane-wave of the scalar field as an
example, and studying the probability to detect this wave on the space-like
hypersurface of constant proper time $\tau$, we conclude that it is capable of
passing through the center $t=z=0$. The only price paid for this feature is the
full delocalization of the state along the hyperplanes $\tau^2=t^2-z^2=0$.  The
state is completely delocalized at $\tau p_t\ll 1$, and it is sharply localized
in the rapidity direction at $\tau p_t\gg 1$. In this way, we approach an idea of
{\em wedge dynamics}, which employs the proper time $\tau$ as the natural
direction of the evolution. In Sec.~\ref{subsec:SBe2}, we consider a wave
packet and demonstrate that the process of localization at finite time $\tau$
is physical; it is accompanied by the gradual re-distribution of the charge
density and the current of this charge. From this observation, we may 
anticipate a special role of the magneto-static interactions at the earliest
times, when the process of the charge density  rearrangement is extremely
rapid. Further calculations of paper [IV]  give even more evidence that the
quantum process of delocalization predicted by wedge dynamics is a material
process.

~~(ii)~ As a first step towards practical calculations, the fields are
described classically and quantized in the scope of wedge dynamics.  In
Sec.~\ref{sec:SN4} of this paper, we accomplish this procedure for the fermion
fields. In Sec.~\ref{sec:SN5}, we derive the expressions for various quantum
correlators, which are used for the  perturbative calculation of the fermion
self-energy in  paper [IV]. An important observation made at this point is that
the material parts of the field correlators immediately have the form of Wigner
distributions. This is a unique property of wedge dynamics which relies on  the
highly localized states as its one-particle basis.

~~(iii)~ The third one, technically the most complicated paper [III] of this
cycle, is dedicated to the vector gauge field in wedge dynamics. Several
conceptual and technical problems are addressed there.  First of all, the
states of the free radiation field are studied classically. Also at the
classical level, we compute the retarded Green function of the vector gauge
field and explicitly separate the longitudinal  (i.e., governed by Gauss law)
field and the field of radiation. It is found that if the physical charge
density $\rho=\tau j_\tau$ vanishes at the starting point $\tau=0$, then 
Gauss law of the wedge dynamics, being in fact a constraint, becomes an
immediate consequence of the equations of motion. Therefore, Gauss law can
be explicitly used to eliminate the unphysical degrees of freedom of the gauge
field, and the gauge $A^\tau=0$ can be fixed completely. Using this result, we
were able to quantize the gluon field according to the standard procedure of
canonical quantization.

The requirement $\rho(\tau=0)=0$ would not be physical in QED, where the
long-range proper fields of electric charges would limit the possible
localization of the first interaction, and the applicability of the wedge
dynamics. On the other hand, in the wedge dynamics of colorless objects built
from the colored fields, which are ``stretched'' at $\tau\to 0$ along a very
wide rapidity interval, this can be a true initial condition. The later
creation of the localized color charges can indeed be initiated by the color
currents in the color-neutral (at $\tau=0$) system.

~~(iv)~ A distinctive property of the longitudinal gauge fields in wedge
dynamics is that they do not look like usual static fields. The Hamiltonian
time $\tau$ does not coincide with a usual time of some particular inertial
Lorentz frame. This is a proper time for all observers that move with all
possible rapidities starting from the point $t=z=0$. The system, which is static
with respect to this time experiences a permanent expansion, and its Gauss
fields have magnetic components. As a consequence, the longitudinal part of the
gauge field propagator acquires a contact term,
$$D^{[contact]}_{\eta\eta}=-{\tau_1^2-\tau_2^2 \over 2} \delta (\eta) \delta
(\vec{r_t})~.$$ The component $D_{\eta\eta}$  establishes a connection between
the $A_\eta$ component of the potential and the $j_\eta$ component of the
current. In its turn, $A_\eta$ is responsible for the $\eta$-component
$E_\eta=\partial_\tau A_\eta$ of the electric field and the $x$- and
$y$-components, $B_x=\partial_y A_\eta$,  $B_y=-\partial_x A_\eta$ of the
magnetic field. The electrical field in the longitudinal $\eta$-direction is
not capable of producing the scattering with transverse momentum transfer.
However, this transfer can be provided by the magnetic forces; the two currents
$j_\eta$ can interact via the magnetic field ${\vec B}_t=(B_x,B_y)$. The origin
of these currents is intrinsically connected with the geometry of states in the
wedge form of dynamics. The existence of these currents  indicates that the
delocalization of the nuclear wave packet is more than a formal decomposition
in terms of fancy modes. This is a physical phenomenon which plays an important
role in the formation of the IR scale of the entire process.

\subsection{Calculation of the effective mass}
\label{subsec:So4} 

The first calculation that incorporates both the ideas and technical part of
the wedge dynamics is attempted in paper [IV]. We compute the effective
``transverse mass'' $\mu(\tau,p_t)$ of the soft  (i.e., $\tau p_t<1$) quark 
mode propagating through the expanding background of hard (i.e., $\tau k_t>1$)
partons. 

~~(i)~In order to find the normal modes of the quark field in the expanding
quark-gluon system, we solve the Dirac equation with radiative
corrections, which can be derived as a projection of the Schwinger-Dyson
equation for the retarded quark propagator onto the one-particle initial
state. This equation can be converted into a dispersion equation that includes
the retarded self-energy and  connects the effective transverse mass
$\mu(\tau,p_t)$ of the soft mode with its transverse momentum $p_t$. This
equation depends on the current proper time $\tau$ as a parameter,
$$\mu(\tau,p_t) = p_t +  \int_{0}^{\tau}d\tau_2 \sqrt{\tau\tau_2} 
e^{i\mu(\tau,p_t)(\tau-\tau_2)}  \Sigma_{ret}(\tau,\tau_2;p_t)~,$$
and we assumed that $d \ln\mu/d\ln\tau\ll 1$, in deriving it. The solution
with this property is indeed found.

~~(ii)~  The material part of the self-energy can be divided into several
parts corresponding to different processes of the forward quark scattering on
the hard partons of the expanding surroundings. First, the quark may scatter on
a real (transverse) gluon. The second process is quark-quark scattering,
which can be conveniently divided into two subprocesses. In one of them, the
interaction is mediated by the radiation part of the gluon field, in the other,
the mediator is the longitudinal part. The latter can be split further into the
contact and non-local parts. Our strategy was to find the leading terms of the
self-energy which are singular at $\tau-\tau_2=0$, and thus can significantly 
contribute to the effective quark mass within a short time. Indeed, since we are
looking for the time-dependent $\mu(\tau,p_t)$, this mass has to be formed
during a sufficiently short time interval. Accordingly, we have chosen the
dimensionless parameter $\xi=(\tau-\tau_2)/\sqrt{\tau\tau_2}$ as a small
parameter.

~~(iii)~ The distribution of hard quarks and gluons that may provide an
effective mass to a soft quark mode with transverse momentum $p_t$ at
the time $\tau\leq 1/p_t$ are taken in agreement with the qualitative
arguments of sections \ref{subsec:So1} and \ref{subsec:So2},
$$ n_f(q_t,\theta)\approx {{\cal N}_f\over \pi R_\bot^2}
{\theta (q_t-p_\ast) \over q_t^2},~~~
 n_g(k_t,\alpha)\approx {{\cal N}_g\over \pi R_\bot^2}
 {\theta (k_t-p_\ast) \over k_t^2}~.$$
They are not related to any dynamical scale and the normalization factors
${\cal N}_g$ and ${\cal N}_f$ are the only (apart from the coupling
$\alpha_s$) parameters of the theory. The impact cross section $\pi
R_\bot^2$ and the full width  $2Y$ of the rapidity plateau are defined by
the geometry of a particular collision and the c.m.s. energy, respectively.
These  are irrelevant for the local screening parameters we are interested in.

~~(iv)~ Analysis of the terms that include radiation fields clearly reveals
two trends. On the one hand, the integration over the transverse momenta of
hard quarks and gluons is capable of creating a singularity when the rapidities
corresponding to the two lines in the loop coincide.  On the other hand, the
interval of rapidities where the collinear geometry is possible is extremely
narrow due to the light-cone boundaries (causality) of the forward scattering
process. The second factor always wins, and the contribution of the collinear
domain is always small. We also found that the observed intermediate  collinear
enhancement of the forward scattering amplitude is, as a matter of fact,
fictitious. It is entirely formed by the integration over the infinitely large
transverse momenta which are physically absent in the distribution of the hard
partons. (Formally, the infinite transverse momentum is needed to provide a
precise tuning of two states with given rapidities to each other.) These
collinear singularities are integrable, and they do not lead to a disaster of
collinear divergence. 

Our  way to pick out the leading contributions  from the space-time domains,
where the phases of the interacting fields are stationary, is a generalization
of the known method of isolating the leading terms using the pinch-poles in the
plane of complex energy. The wedge dynamics does not allow for a standard
momentum representation, since its geometric background is not homogeneous in
$t$- and $z$-directions. Nevertheless, the patches of phase space, where the
phases of certain field fragments are stationary and effectively overlap, do
now the same job as the pinch-poles, and yield the same answers when the
homogeneity required for the momentum representation is restored. This way to
tackle the problem is genuinely more general, because it addresses the
space-time picture of the interacting fields. The role of pinch-poles is taken
over by the geometrical overlap of the field patterns with the same rapidity.
This observation can serve as a footing for the future development of  an
effective technique for perturbative calculations in wedge dynamics.

~~(v)~ The effect of the non-local components of the longitudinal part of the 
gluon propagator that mediates the quark-quark scattering, was shown to be
small also. This interaction cannot lead to the collinear enhancement. However,
its yield could be not very small, because the interaction has long range. It
occurs, that the non-local electro- and magneto-static interactions of charges 
just almost compensate each other.

~~(vi)~ The only term in the quark self-energy which is singular at small time
differences is due to the above mentioned contact term in the $D^{\eta\eta}$
component of the gluon propagator. This is the leading contribution to the
dispersion equation  provided by the magneto-static interaction of the
longitudinal currents. Studied in the first approximation, the solution of the
dispersion  equation indicates that in compliance with the original idea, the
effective mass $\mu(\tau,p_t)$ gradually increases with time reaching its
maximum when $\tau p_t\approx 1$. This is the major practical result of this
study. Evolution of the fields at the later times must be approached with
another set of normal modes that, from the very beginning, account for the
screening effects developed at the previous stage.

\subsection{ Conclusion and perspectives. }
\label{subsec:So5} 

In a series of papers reviewed in this section, we demonstrated that the field
theory is indeed able to describe a scenario. By scenario we mean a continuous
smoothly developing temporal sequence of one stage into another. These stages
are different only in the respect that each of them is characterized by its
individual {\em optimal set} of  normal modes. In contrast with paper [I], we
do not focus on the stage of QCD evolution, since we have no clear image of the
objects that initiate destruction of nuclear coherence. Instead, we try to
understand, what can be the immediate products of this destruction.  This is an
example of the continuity that stands behind the idea of the scenario. The next
stage will be the kinetics of the partons-plasmons, and we anticipate that it
will impose new restrictions, which will improve our current results. [Quantum
mechanics works remarkably in both directions: any information about the
properties of the final state imposes limitations on the possible line of the
evolution (including the initial data) at the earlier times exactly in the same
way as the known initial data imposes restrictions on the possible final
states.] By the same token, we must look for a connection between the objects
resolved in the first interaction of two nuclei and the known properties of
hadrons and the QCD vacuum. Unfortunately, this appealing opportunity is still
distant. 

First principles appeared to be a powerful tool for achieving our goals.
With minimal theoretical input and with the reference to the simplest data,
they allow one to build a self-consistent picture of the initial stage of
the collision. Colliding the nuclei, we probably create the theoretically
simplest situation  for understanding the nature of the process. In the
course of this study, we relied only on the fact of boost invariance of the
process  and an assumption that the field states with large transverse
momentum, even at very early times, may be associated with the localized
particles and thus can be described by the distribution with respect to their
rapidity and transverse momentum.  Our strategy of looking for the leading
contributions and all our approximations in calculating the material part
of the quark self-energy are based on this assumption. If it appears
incorrect, then  it is most likely that the quark-gluon matter created in
the collision of  two nuclei never, and in no approximation, can be
considered as a system of nearly free and weakly interacting field states. 

Our decision to begin the exploration of potentialities of the wedge dynamics
with the  computation of quark self-energy is motivated only by technical
reasons. The gluon propagator of wedge dynamics is a very complicated function,
and we preferred to start with the computation of the fermion loop which has
only one gluon correlator in it. We hope that the discovery of, in the course
of our study, an enormous simplifications (with respect to what we had to start
with) will allow us to address the more important problem of the gluon
self-energy in a reasonably economic way.

\section{Field states in the proper-time dynamics}
\label{sec:S2}

The dynamical masses of normal modes at finite density are found from the
dispersion equation that includes the corresponding self-energy, i.e., the
amplitude of the forward scattering of the mode on the particles that populate
the phase space.  In paper [I], we found that it is impossible to adequately
describe this basic process of  forward scattering in the null-plane
dynamics. The problem arises due to the singular behavior of the field
pattern which is defined as the static field with respect to the Hamiltonian
time $x^+$. This singular behavior alone shows that the  choice of the
dynamics
and the proper definition of the  field states is a highly  nontrivial and
important issue. Besides, if we tried to describe quantum fluctuations in the
second nucleus in the same fashion, then it would require a second
Hamiltonian time $x^-$, which is not acceptable.  Thus,  if we wish to view
the collision of two nuclei as a unique quantum process, then it is
imperative to find a  way to describe quarks and gluons of both nuclei,  as
well as the products of their interaction, using {\it the same Hamiltonian
dynamics}.  An appropriate choice for the gluons is always difficult because
the  gauge  is a global object (as are the Hamiltonian dynamics) and both
nuclei should be described using the same  gauge condition.

Quantum field theory has a strict definition of  {\it dynamics}. This
notion was introduced by Dirac  \cite{Dirac}  at the end of the 1940's in
connection with his attempt to build a quantum theory of the gravitational
field. Every (Hamiltonian) dynamics includes its specific definition of
the quantum mechanical observables on the (arbitrary) space-like surfaces,
as well as the means to describe the evolution of the observables  from
the ``earlier'' space-like surface to the ``later'' one. 

The primary choice of the degrees of freedom  is effective if, even without
any interaction, the dynamics of the normal modes adequately  reflects the
main physical features of the phenomenon.  The intuitive physical arguments
clearly indicate that the normal modes of the fields participating in the
collision of two nuclei should be compatible with their Lorentz
contraction. Unlike the incoming plane waves of the standard scattering
theory, the nuclei have a well-defined shape and the space-time domain of
their intersection is also well-defined. Hence, the geometric properties of
the expected normal modes follows, in fact, from the uncertainty principle.
Indeed, we may view the first touch of the nuclei as  the first of the two
measurements which are necessary to determine the velocity.  Since a
precise measurement of the nuclei coordinate at an exactly determined
moment appears to be an inelastic process that completely destroys the
nuclei, the spectrum of the longitudinal velocities of the
final-state components must become extremely wide \cite{Landau}.  These
components may also be different by their transverse momenta. With respect
to the measurement of the longitudinal velocity, the latter plays a role of
an ``adjoint mass''. The velocity of a heavier object can be measured with
a larger accuracy. Therefore, the separation of the ``heavy'' final state
fragments by their longitudinal velocities requires less time and can be
verified earlier than for the ``light'' ones.\footnote{The boost-invariance
with the fixed center means the absence of a corresponding scale and {\em
vice-versa}. Any relativistic equations, regardless of their physical
content, will yield a self-similar solution. For example, the relativistic
hydrodynamic equations lead to a known Bjorken solution with the rapidity
plateau. In its turn, the Bjorken solution can be obtained as a limit of
the Landau solution with an infinite Lorentz contraction of the colliding
objects. We favor the arguments that are closer to quantum mechanics and
allow for the further connection with the properties of the quantum states.}

The same conclusion can be reached formally: Of the ten symmetries of the
Poincar\'{e} group, only  rotation around the  collision $z$-axis, boost along
it,  and the translations in the transverse $x-$ and $y$-directions 
survive. The idea of the collision  of two plane sheets immediately leads us
to  the {\em wedge form};  the states of quark and gluon fields before  and
after the collision must be confined within the past and future  light cones
(wedges) with  the $xy$-collision plane as the edge. Therefore, it is
profitable to choose, in advance, the set of normal modes which have the
symmetry of the localized initial interaction and carry quantum numbers
adequate to this symmetry. These quantum numbers are the transverse components
of momentum and the  rapidity of the particle (which replaces the component
$p^z$ of its momentum). In this {\it ad hoc} approach, all the spectral
components of the nuclear  wave functions ought to collapse in the
two-dimensional plane of the interaction,  even if all the confining
interactions of the quarks and gluons in the hadrons  and the coherence of the
hadronic wave functions are neglected. 

In the wedge form of dynamics, the states of  free quark and gluon fields are
defined (normalized) on the space-like hyper-surfaces of the constant proper
time $\tau$, $\tau^2=t^2-z^2$. The main idea of this approach is to study the
dynamical evolution of the interacting fields along the Hamiltonian time
$\tau$. The gauge of the gluon field is fixed by the condition $A^\tau =0$. 
This simple idea solves several problems. On the one hand,   it becomes
possible to treat the two different light-front dynamics  which describe  each
nucleus of the initial state separately, as  two limits of this single
dynamics.  On the other hand, after the collision, this  gauge simulates a 
local (in rapidity) temporal-axial gauge. This feature provides a smooth
transition to the boost-invariant regime of the  created matter expansion (as a
first approximation).  Particularly, addressing the problem of screening, we
will be able to compute the plasmon mass in a uniform fashion, considering each
rapidity interval separately.

As it was explained in the first two sections, the feature of the states to
collapse at the interaction vertex  is crucial for understanding the dynamics
of a high-energy nuclear collision. A simple optical prototype of the wedge
dynamics is the {\em camera obscura} (a dark chamber with the pin-hole in the
wall). Amongst the many possible {\em a priori} ways to decompose the incoming
light, the camera selects  only one. Only  the spherical harmonics centered at
the pin-hole can penetrate  inside the camera.  The spherical waves reveal
their angular dependence  at some distance from the center and build up the
image on the opposite wall. Here, we suggest to view the collision of two
nuclei as a kind of diffraction of the initial wave functions through the
``pin-hole'' $t=0,~z=0$ in $tz$-plane.

Using the proper time $\tau$ as the natural direction of the evolution of the 
nuclear matter after the collisions has far reaching consequences. The
surfaces of constant $\tau$ are curved, and the oriented objects like spinors
and vectors have to be defined with due respect to this curvature. We have to
incorporate the tetrad formalism in order to differentiate them covariantly.
The properties of local invariance are modified also, since the different
directions in the tangent plane become not equivalent. The physical content of
the theory also undergoes an important change. The system of observers that
are used to  {\em normalize} the quantum states of wedge dynamics is different
from the observers of any particular inertial Lorentz frame.

\subsection{One-particle wave functions in wedge dynamics}
\label{subsec:SBe1} 

In order to study the main kinematic properties of the states of the wedge
dynamics, it is enough to consider the one-particle wave functions  of the
scalar field. Let us take the wave function $\psi_{\theta,p_\perp}(x)$ of the
simplest form,
\begin{eqnarray}
\psi_{\theta,p_\perp}(x)= {1 \over 4\pi^{3/2}} 
e^{-ip^0t+ip^zz+i{\vec p}_{\perp}{\vec r}_{\perp}}\equiv
\left\{ \begin{array}{l}  4^{-1}\pi^{-3/2}
e^{-im_{\perp}\tau\cosh(\eta-\theta)}
e^{i{\vec p}_{\perp}{\vec r}_{\perp}},~~~~\tau^2>0, \\
 4^{-1}\pi^{-3/2}  e^{-im_{\perp}\tau\sinh(\eta-\theta)}
e^{i{\vec p}_{\perp}{\vec r}_{\perp}},~~~~ \tau^2<0~. \end{array}\right.
\label{eq:S1.1}\end{eqnarray} 
where $p^0=m_{\perp}\cosh\theta$, $p^z=m_{\perp}\sinh\theta$ ($\theta$ being 
the rapidity of the particle), and, as usual,
$m_{\perp}^{2}=p_{\perp}^{2}+m^2$. The above form implies that $\tau$ is
positive in the future of the wedge vertex  and negative in its past. Even
though this wave function is obviously a plane wave which occupies the whole
space, it carries the  quantum number $\theta$ (rapidity of the particle)
instead of the momentum $p_z$. A peculiar property of this wave function is
that it may be normalized in two different ways, either on the hypersurface
where $t=const$,
\begin{eqnarray}
\int_{t=const}\psi^{\ast}_{\theta',p'_\perp}(x)
~i{\stackrel{\leftrightarrow}{\partial \over \partial t}}~ 
\psi_{\theta,p_\perp}(x)~dz d^2{\vec r}_{\bot}=
\delta (\theta -\theta')\delta ({\vec p}_{\bot}-{\vec p'}_{\bot})~~,
\label{eq:S1.2}\end{eqnarray}   
or, equivalently, on the hypersurfaces $\tau=const$ in the 
future- and the past--light wedges of the collision plane, where $\tau^2>0$,
\begin{eqnarray}
\int_{\tau=const}\psi^{\ast}_{\theta',p'_\perp}(x)
~i{\stackrel{\leftrightarrow}{\partial \over \partial \tau}}~ 
\psi_{\theta,p_\perp}(x)~\tau d\eta d^2{\vec r}_{\bot}=
\delta (\theta -\theta')\delta ({\vec p}_{\bot}-{\vec p'}_{\bot})~.
\label{eq:S1.3}\end{eqnarray} 
 Being almost identical mathematically, these two equations are very
different physically. Eq.~(\ref{eq:S1.2}) implies that the state is
detected by a particular Lorentz observer equipped by a grid of detectors
that cover the whole space, while Eq.~(\ref{eq:S1.3})  normalizes the
measurements performed by an array of the detectors moving with all
possible velocities. At any particular time of the Lorentz observer, this
array even does not cover the whole space.

The norm of a particle's wave function always corresponds to the conservation
of its charge or the probability to find it. Since the norm given by
Eq.~(\ref{eq:S1.3}) does not depend on $\tau$, the particle with a given
rapidity $\theta$ (or velocity $v=\tanh\theta=p^z/p^0$), which is
``prepared''  on the surface  $\tau=const$ in the past light wedge, cannot
flow through the light-like wedge boundaries; the particle is
predetermined to penetrate in the future light wedge through its vertex. The
dynamics of the penetration process can be understood in the following way.

At large $~m_{\perp}|\tau|$, the phase of the wave function 
$\psi_{\theta,p_\perp}$ is stationary in a very narrow interval  around
$~\eta=\theta~$ (outside this interval, the function oscillates with
exponentially increasing  frequency); the wave function describes a particle
with rapidity $~\theta$ moving along the classical trajectory.  However, for
$m_{\perp}|\tau|\ll 1$, the phase  is almost constant along the surface
$\tau=const$.  The smaller $\tau~$ is, the more uniformly the domain of 
stationary  phase is stretched out along the light cone. A single particle
with the wave function $\psi_{\theta,p_\perp}$ begins its life as the wave 
with the given rapidity $\theta$ at large negative $\tau$. Later, it becomes
spread out over the boundary of the past light wedge as $\tau\to -0$. Still
being spread out, it appears on the boundary of the future light wedge.
Eventually, it again becomes a wave  with  rapidity $\theta$ at large positive
$\tau$.  The size and location of the interval where the phase of the wave
function is stationary plays a central role in all subsequent discussions,
since it is equivalent to the  localization of a particle. Indeed, the
overlapping of the domains of stationary phases in space and time provides the
most  effective interaction of the fields.

The size $\Delta\eta$ of the $\eta$-interval  around the particle rapidity
$\theta$, where the wave function is  stationary, is easily evaluated.
Extracting  the trivial factor $e^{-im_{\perp}\tau}$ which defines the
evolution of the wave  function in the $\tau$-direction, we obtain an
estimate from the exponential of  Eq.~(\ref{eq:S1.1}),
\begin{eqnarray}
 2~m_{\perp}\tau\sinh^2(\Delta\eta/2)\sim 1~~. 
\label{eq:S1.4}\end{eqnarray}      
The two limiting cases are as follows, 
 \begin{eqnarray}
 \Delta\eta \sim \sqrt{2\over m_{\perp}\tau},
~~{\rm when}~~ m_{\perp}\tau \gg 1~,~~~{\rm and}~~~
  \Delta\eta \sim 2 \ln {2\over m_{\perp}\tau},
~~{\rm when}~~ m_{\perp}\tau\ll 1~~.
\label{eq:S1.5}\end{eqnarray} 
In the first case, one may boost this interval into the laboratory reference
frame and see that the interval of the stationary phase is Lorentz  contracted
(according to the rapidity $\theta$) in $z$-direction.  This estimate
confirms what follows from physical intuition; for a heavier quantum object,
the velocity can be measured with the higher accuracy. The states of the wedge
dynamics appear to be almost ideally suited for the analysis of the processes
that are localized at different times $\tau$ and intervals of rapidity $\eta$,
and are characterized by a different transverse momentum transfer.  With
respect to any particular process, these states are  easily divided into 
slowly varying fields and localized particles. In this way, one may introduce
the distribution of particles and study their effect on the dynamics of the
fields.  As a result, we can calculate  the plasmon mass as a local (at some
scale) effect which agrees with our understanding of its physical origin.

\subsection{Dynamical properties of states in wedge dynamics}
\label{subsec:SBe2} 

The property of the wave function to concentrate with the time near the
classical world line of a particle with the given velocity has important
implications. This is a gradual process and it must be accompanied by the
re-distribution of the charge density and the current of this charge. To see
how this happens explicitly, let us consider a particle in a superposition
state  $|\theta_0\rangle$ of a normalized wave packet,
\begin{eqnarray}
 |\theta_0\rangle =\int_{-\infty}^{\infty}d\theta f(\theta-\theta_0)
 \bbox{a^\dag}_\theta |0\rangle~,~~~~~
 \langle \theta_0|\theta_0\rangle = \int_{-\infty}^{\infty}d\theta 
 f^\ast(\theta-\theta_0) f(\theta-\theta_0) = 1~,
\label{eq:S1.6}\end{eqnarray} 
where $\bbox{a^\dag}_\theta$ is the Fock creation operator for the
one-particle state with the rapidity $\theta$.\footnote{We do not describe
here the procedure of the scalar field quantization in wedge dynamics. It is
exactly the same as quantization of the fermion field in the next section.} 
The explicit form of the weight function in Eq.~(\ref{eq:S1.6}) may vary.
Solely for convenience, we take the weight function $f(\theta-\theta_0)$ of
the form,
\begin{eqnarray}
 f(\theta-\theta_0) = [K_{0}(2\xi)]^{-1/2} e^{-\xi\cosh(\theta-\theta_0)}
 \approx (4\xi /\pi)^{1/4} e^\xi~e^{-\xi\cosh(\theta-\theta_0)}~,
\label{eq:S1.7}\end{eqnarray} 
which provides a sharp localization of the wave packet. In the second of
these equations, we used an asymptotic approximation of the Kelvin function
$K_0(2\xi)$, which is reasonably accurate starting from $\xi\geq 1/2$.

\begin{figure}[htb]
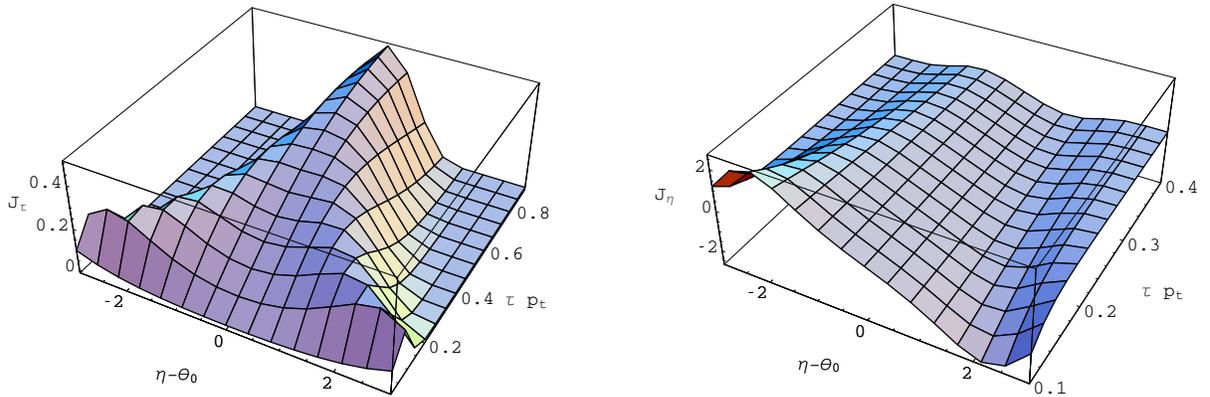

\begin{center}
\mbox{ 
\psfig{file=./fig1b.ps,height=2.2in,bb=100 490 390 710}
\hspace{1.cm}
\psfig{file=./fig1a.ps,height=2.2in,bb=100 490 390 710}
}
\end{center}
\caption{Charge density in the wave packet (left) and current density
(right) evolution.}
\label{fig:fig1}
\end{figure}

The operator of the four-current density for the complex scalar field
$\Psi$ is well known to be
\begin{eqnarray}
J_\mu (x) = \Psi^\dag (x)~
i{\stackrel{\leftrightarrow}\partial_\mu}~\Psi (x)~,
\label{eq:S1.8}\end{eqnarray} 
and to obey the covariant conservation law,
\begin{eqnarray}
\nabla_\mu J^\mu (x) = 
(-g)^{-1/2}\partial_\mu [(-g)^{1/2} g^{\mu\nu}J_\nu (x)]
= \tau^{-1}[\partial_\tau(\tau J_\tau) + \partial_\eta(\tau^{-1}J_\eta)]=0~.
\label{eq:S1.9}\end{eqnarray}
(Here, for simplicity,  we consider the two-dimensional case and employ the 
metric  $g^{\tau\tau}=1$, $g^{\eta\eta}=-\tau^{-2}$.) The physical
components of the current (which are defined in such a way that the integral
form of the conservation law is not altered by the curvilinear metric) are
${\cal J}_\tau=\tau J_\tau $ and ${\cal J}_\eta=\tau^{-1}J_\eta$. Using
Eqs.~(\ref{eq:S1.6}) and (\ref{eq:S1.8}), we can compute their expectation
values of these components in the state $|\theta_0\rangle$.
\begin{eqnarray}
\langle \theta_0|{\cal J}_\tau|\theta_0\rangle = \tau k_t
\int_{-\infty}^{\infty}{d\theta_1d\theta_2\over 4\pi}
f^\ast(\theta_1-\theta_0)f(\theta_2-\theta_0)
[\cosh(\eta -\theta_1)+\cosh(\eta -\theta_2)]
e^{i\tau k_t[\cosh(\eta -\theta_1)-\cosh(\eta -\theta_2)]}~,
\label{eq:S1.10}\end{eqnarray}
\begin{eqnarray}
\langle \theta_0|{\cal J}_\eta|\theta_0\rangle = {k_t\over\tau}
\int_{-\infty}^{\infty}{d\theta_1d\theta_2\over 4\pi}
f^\ast(\theta_1-\theta_0)f(\theta_2-\theta_0)
[\sinh(\eta -\theta_1)+\sinh(\eta -\theta_2)]
e^{i\tau k_t[\cosh(\eta -\theta_1)-\cosh(\eta -\theta_2)]}~.
\label{eq:S1.11}\end{eqnarray}
The integrals over $\theta_1$ and $\theta_2$ can be estimated by means of
the saddle point approximation even for the relatively small values of $\xi$,
e.g. $\xi\sim 1$, because the hyperbolic functions in the exponents vary
sufficiently rapidly near the stationary points. These calculations yield the
following result,
\begin{eqnarray}
\langle \theta_0|{\cal J}_\tau|\theta_0\rangle = 
{2\tau k_t\over (\pi\xi)^{1/2}}~{\cosh(\eta-\theta_0)\over
1+{\tau^2k_t^2\over\xi^2}\cosh[2(\eta-\theta_0)]}~
{e^{-{\tau^2k_t^2\over\xi}\sinh^2(\eta-\theta_0)}\over
\sqrt{1+{\tau^2k_t^2\over\xi^2}\sinh^2(\eta-\theta_0)}}  ~,
\label{eq:S1.12}\end{eqnarray}
\begin{eqnarray}
\langle \theta_0|{\cal J}_\eta|\theta_0\rangle = 
{2k_t\over (\pi\xi)^{1/2}}~{\sinh(\eta-\theta_0)\over
1+{\tau^2k_t^2\over\xi^2}\cosh[2(\eta-\theta_0)]}~
{e^{-{\tau^2k_t^2\over\xi}\sinh^2(\eta-\theta_0)}\over
\sqrt{1+{\tau^2k_t^2\over\xi^2}\sinh^2(\eta-\theta_0)}} ~.
\label{eq:S1.13}\end{eqnarray}
These dependences are plotted in Fig.~\ref{fig:fig1} up to a common
scale factor.
\bigskip

{}From the left figure, it is easy to see that the evolution of the charge
density ${\cal J}_\tau$ starts from the lowest magnitude and the widest spread
at small $\tau$. Then it gradually becomes narrow and builds up a significant
amplitude near the classical trajectory with the rapidity $\theta_0$. This
process is accompanied by the  charge flow ${\cal J}_\eta$ (right figure) which
has its maximal values at small $\tau$, and then gradually vanishes at later
times, when the process of building the classical particle comes to its
saturation. The extra factor $\tau^{-1}~$ in $~\langle \theta_0|{\cal J}_\eta
|\theta_0\rangle$, which tends to boost current at small $\tau$, is of
geometric origin. Thus, the behavior of the local observables in the wave
packet confirms the simple arguments of Sec.~\ref{subsec:SBe1} based on the
analysis of the domain where the wave function is stationary.  One can guess
about the possible nature of interactions at the earliest times by making an
observation that the $\eta$-component of the current must produce the $x$- and
$y$-components of the magnetic field. These fields are the strongest at the
earliest times when $\tau k_t\ll 1$.  The magnetic fields of the transition
currents provide scattering with the most effective transfer of the transverse
momentum. Indeed, at time $\tau_2$ the quark with the transverse momentum
$p_t$, $\tau_2 p_t\ll 1$ interacts with the gluon field and acquires a large
transverse momentum $k_t$, $\tau_2 k_t\gg 1$. This transition is characterized
by a drastic narrowing of the charge spread in the rapidity direction, and must
be accompanied by a strong $\eta$-component of the transition current. A
similar transition in the opposite direction happens at the time $\tau_1$, when
the gluon field interacts with another quark that has large initial transverse
momentum $k_t$, and recovers the soft state with $\tau_2 p_t\ll 1$ in the
course of this interaction. This second transition current readily interacts
with the magnetic component of the gluon field. These speculations will be
justified in paper [IV] of this cycle  by the explicit calculation of quark
self-energy in the expanding system.  

Three remarks are in order: First, the field of a free on-mass-shell particle
can be only static, and it is common to think that, in the rest frame of the
particle, it is a purely electric field. In the wedge dynamics, the particle is
formed during a finite time and this formation process unavoidably generates
the magnetic component of the {\em longitudinal} (i.e., governed by the Gauss
law) field. This will be obtained more rigorously in the next paper when the
full propagator of the gauge field in wedge dynamics will be found.
Furthermore, in wedge dynamics, the source must be called as static if it
expands in such a way that ${\cal J}_\tau=\tau J_\tau = const(\tau)$ , and its
field strength also has a magnetic component. Second, the local color current
density may be large even when the system is color-neutral (begins its
evolution from the colorless state), as it seems to be the case in heavy-ion
collisions. It will be also shown in paper [III], that in order to fix the
gauge $~A^\tau=0~$ completely, one must require that the physical charge
density $~{\cal J}^\tau=0~$ at $~\tau=0~$. Finally, in wedge dynamics we meet a
unique structure of phase space, where two variables, the velocity of a
particle and its rapidity coordinate, just duplicate each other at sufficiently
late proper time $\tau$. The quantum mechanical uncertainty principle does not
prohibit one to address them on equal footing, because  the one-particle wave
packets of the wedge dynamics, evolving in time, become more and more narrow in
rapidity direction.

\section{States of Fermions}
\label{sec:SN4}  

The hypersurfaces of constant Hamiltonian time $\tau$ of wedge dynamics are
curved. Therefore, all oriented objects like vectors or spinors are
essentially defined only in the tangent space and therefore, their
covariant derivatives should be calculated in the framework of the
so-called tetrad
formalism \cite{Fock,Witten}.\footnote{In what follows, we use the Greek
indices for  four-dimensional vectors and tensors in the curvilinear
coordinates, and the Latin indices from $a$ to $d$ for the vectors in flat
Minkowsky coordinates. We use Latin indices from $r$ to $w$ for the
transverse $x$- and $y$-components ($r,...,w=1,2$), and the arrows over the
letters to denote the  two-dimensional vectors, {\em e.g.}, ${\vec
k}=(k_x,k_y)$, $|{\vec k}|=k_{t}$. The Latin indices from $i$ to $n$
($i,...,n=1,2,3$) will be used  for the three-dimensional internal
coordinates $u^i=(x,y,\eta)$ on the hyper-surface $\tau=const$.}      
The covariant derivative of the tetrad vector includes two connections
(gauge fields). One of them, the Levi-Civita connection
$$\Gamma^{\lambda}_{~\mu\nu}={1\over 2} {\rm g}^{\lambda\rho}
\bigg[{\partial {\rm g}_{\rho\mu} \over \partial x^\nu }
+{\partial {\rm g}_{\rho\nu} \over \partial x^\mu }
-{\partial {\rm g}_{\mu\nu} \over \partial x^\rho } \bigg]~,$$
is the gauge  field, which provides  covariance with respect to the general
transformation of coordinates. The second gauge field,  the spin
connection  $\omega_{\mu}^{~ab}(x)$, provides covariance with respect to
the local Lorentz rotation. Let $x^\mu =(\tau,x,y,\eta)$ be the
contravariant components of the curvilinear coordinates and
$x^a=(t,x,y,z)\equiv (x^0,x^1,x^2,x^3)$ are those of the flat Minkowsky
space. Then the tetrad vectors $e^{a}_{~\mu}$ can be taken as follows,
\begin{eqnarray}  
e^{0}_{~\mu}=(1,0,0,0),~~e^{1}_{~\mu}=(0,1,0,0),~~
e^{2}_{~\mu}=(0,0,1,0),~~e^{3}_{~\mu}=(0,0,0,\tau)~.
\label{eq:E3.1}\end{eqnarray}        
They correctly reproduce the curvilinear metric ${\rm g}_{\mu\nu}$ and 
the flat Minkowsky metric $g_{ab}$, {\em i.e.},
\begin{eqnarray}
{\rm g}_{\mu\nu}=g_{ab}e^{a}_{~\mu}e^{b}_{~\nu}
={\rm diag}[1,-1,-1,-\tau^2]~,~~~
g^{ab}= {\rm g}^{\mu\nu} e^{a}_{~\mu}e^{b}_{~\nu}={\rm diag}[1,-1,-1,-1] ~.
\label{eq:E3.2}\end{eqnarray} 
The spin connection can be found from the condition that the covariant
derivatives of the tetrad vectors are equal to zero,
\begin{eqnarray}
\nabla_\mu e^{a}_{~\nu}=\partial_\mu e^{a}_{~\nu}
+\omega^{~a}_{\mu~b}e^{b}_{~\nu} -
\Gamma^{\lambda}_{~\mu\nu}e^{a}_{~\lambda} =0~.
\label{eq:E3.3} 
\end{eqnarray} 
The covariant derivative of the spinor field includes only the spin
connection,
 \begin{eqnarray}  
\nabla_\mu\psi(x)= \big[\partial_\mu +
{1\over 4}\omega_{\mu}^{~ab}(x)\Sigma_{ab}\big]\psi(x)~,
\label{eq:E3.4}\end{eqnarray} 
where $\Sigma^{ab}={1\over 2}[\gamma^a\gamma^b-\gamma^b\gamma^a]$ is
an obvious generator of the Lorentz rotations and $\gamma^a$ are the
Dirac matrices of Minkowsky space. Introducing the Dirac matrices
in curvilinear coordinates, $\gamma^{\mu}(x)=e^{\mu}_{~a}(x)\gamma^a$, 
one obtains the Dirac equation in curvilinear coordinates,       
\begin{eqnarray}  
[\gamma^\mu(x)(i\nabla_\mu +gA_\mu(x))-m]\psi(x)=0 ~,
\label{eq:E3.5}\end{eqnarray}   
where $A^\mu(x)$ is the gauge field associated with the local group of 
the internal
symmetry. The conjugated spinor is defined as
${~\overline \psi}=\psi^{\dag}\gamma^0$, and obeys the equation, 
\begin{eqnarray}  
(-i\nabla_\mu +gA_\mu(x)){\overline\psi}(x)\gamma^\mu(x)- 
m{\overline\psi}(x)=0 ~~.
\label{eq:E3.6}\end{eqnarray}  
 
These two Dirac equations correspond to the action, 
\begin{eqnarray}
{\cal A}=\int d^4 x \sqrt{-g}{\cal L}(x)=
\int d^4 x \sqrt{-g} \{ {i\over 2}[{\overline\psi}\gamma^\mu(x)
\nabla_\mu\psi- (\nabla_\mu {\overline\psi})\gamma^\mu(x)\psi]
+ g{\overline\psi}\gamma^\mu(x) A_\mu\psi -
m{\overline\psi}\psi \}~~,
\label{eq:E3.7}    
\end{eqnarray} 
from which one  easily obtains the locally conserved $U(1)$-current,         
\begin{eqnarray}
J^\mu(x)= {\overline\psi}(x)\gamma^\mu(x)\psi(x)~, 
~~~~ (-{\rm g})^{-1/2}\partial_\mu [(-{\rm g})^{1/2}
{\rm g}^{\mu\nu}(x)J_\nu (x)]=0~.
\label{eq:E3.8}\end{eqnarray}  
The Dirac equations (\ref{eq:E3.5}) and (\ref{eq:E3.6}) can be 
alternatively obtained as the equations of the
Hamiltonian dynamics along the proper time $\tau$.
The canonical momenta conjugated to the fields $\psi$ and ${\overline\psi}$
are                                 
\begin{eqnarray} 
\pi_{\psi}(x)={\delta(\sqrt{-{\rm g}}{\cal L})\over\delta \dot{\psi}(x)}
= {i\tau\over 2} {\overline\psi}(x)\gamma^0
\;\; {\rm and} \;\;\;
\pi_{{\overline\psi}}(x)=
{\delta(\sqrt{-{\rm g}}{\cal L})\over\delta \dot{{\overline\psi}(x)} }
= -{i\tau\over 2} \gamma^0\psi(x)~,
\label{eq:E3.9}\end{eqnarray}  
 respectively.
The Hamiltonian of the Dirac field in the wedge dynamics has the following
form,
\begin{eqnarray} 
H=\int d\eta d^2 {\vec r}\;\sqrt{-{\rm g}} 
\{ -{i\over 2}[{\overline\psi}\gamma^i(x)
\nabla_i\psi- (\nabla_i {\overline\psi})\gamma^i (x)\psi]
- g{\overline\psi}\gamma^\mu(x) A_\mu\psi+
m{\overline\psi}\psi \}~,
\label{eq:E3.10}\end{eqnarray}  
 and the wave equations are just the Hamiltonian equations of
motion for the momenta.                                           
 
The non-vanishing components of the connections are 
$\Gamma^{\bf\cdot}_{\eta\tau\eta} = \Gamma^{\bf\cdot}_{\eta\eta\tau}=
-\Gamma^{\bf\cdot}_{\tau\eta\eta}=-\tau$~ and ~$\omega_{\eta}^{~30}
=-\omega_{\eta}^{~03}=1$. Moreover, we have $\gamma^\tau(x)=\gamma^0$
and  $\gamma^\eta(x)=\tau^{-1}\gamma^3 $. The explicit form of the
Dirac equation in our case is as follows,
\begin{eqnarray}  
[i \not\!\nabla -m]\psi(x)= [i\gamma^0(\partial_\tau +{1\over 2\tau})
+i\gamma^3{1\over \tau}\partial_\eta +i \gamma^r \partial_r -m]\psi(x)=0~~.
\label{eq:E3.11}\end{eqnarray} 
The one-particle solutions of this equation must be normalized according to
the
charge conservation law (\ref{eq:E3.8}). We choose  the  scalar
product  of the following form,
\begin{eqnarray}  
(\psi_1,\psi_2)=
\int \tau d\eta ~d^2 {\vec r}~ 
{\overline \psi}_1 (\tau,\eta,{\vec r})\gamma^\tau  
\psi_2 (\tau,\eta,{\vec r})~~.
\label{eq:E3.12}\end{eqnarray}   
With this definition of the scalar product, the Dirac equation is
self-adjoint.
The solutions to this equation will be looked for in the form 
$\psi(x)= [i \not\!\nabla +m]\chi(x)$, with the function $\chi(x)$ that
obeys the ``squared'' Dirac equation,
\begin{eqnarray}  
[i \not\!\nabla -m][-i \not\!\nabla +m]\chi(x)= 
\bigg[\partial_{\tau}^{2} +{1\over \tau}\partial_\tau
-{1\over \tau^2}\partial_{\eta}^{2} - \partial_{r}^{2} +m^2-
\gamma^0\gamma^3 {1\over \tau^2}\partial_{\eta} \bigg]\chi(x)=0~~.
\label{eq:E3.13}\end{eqnarray}       
The spinor part $\beta_\sigma$ of the function $\chi(x)$ can be chosen
as  an eigenfunction of the operator $\gamma^0\gamma^3 $, namely,
$~\gamma^0\gamma^3\beta_{\sigma} = \beta_{\sigma}$, and $\sigma=1,2~$.         
Therefore, the solution of the original Dirac equation 
(\ref{eq:E3.5}) can be written down 
as $~\psi^{\pm}_{\sigma} = w_\sigma \chi^{\pm}(x)$,~   
with the bi-spinor operators $w_\sigma=[i \not\!\nabla +m]\beta_\sigma $
that act on the positive- and negative-frequency solutions $\chi^{\pm}(x)$
of the scalar equation
\begin{eqnarray}  
\bigg[\partial_{\tau}^{2} +{1\over \tau}\partial_\tau
-{1\over \tau^2}(\partial_{\eta}+{1\over 2})^{2} - \partial_{r}^{2} 
+m^2\bigg]\chi^{\pm}(x)=0~~.
\label{eq:E3.14}\end{eqnarray}       
In what follows, we shall employ only the partial waves with quantum numbers
of transverse momentum $\vec{p_t}$ and rapidity $\theta$ of massless quarks. 
In this case, the (already normalized) scalar functions $\chi^{\pm}(x)$ are
\begin{eqnarray}  
\chi^{\pm}_{\theta,{\vec p_t}}(x)=  (2\pi)^{-3/2} (2p_t)^{-1/2} 
~e^{(\theta-\eta)/2}
e^{\mp i p_t\tau\cosh(\theta-\eta)}~e^{\pm i{\vec p}{\vec r}}~.
\label{eq:E3.15}\end{eqnarray} 
Consequently, the one-particle solutions are
\begin{eqnarray}  
\Omega^{\pm}_{\sigma,\theta,{\vec p_t}}(x)=  (2\pi)^{-3/2} (2p_t)^{-1/2} 
i \not\!\nabla~\beta_\sigma e^{(\theta-\eta)/2}
e^{\mp i p_t\tau\cosh(\theta-\eta)}~e^{\pm i{\vec p}{\vec r}}~,
\label{eq:E3.16}\end{eqnarray}
where the spinors $\beta_\sigma$ can be chosen in different ways. However,
regardless of a particular choice of the spinors  $\beta_\sigma$, the
polarization sum is always
\begin{eqnarray}  
\sum_\sigma \beta_\sigma \otimes \beta_\sigma= {1+\gamma^0\gamma^3\over 2}~.
\label{eq:E3.17}\end{eqnarray}  
The waves $\Omega^{\pm}_{\sigma,\theta,{\vec p_t}}(x)$ are orthonormalized
according to
\begin{eqnarray}  
(\Omega^{(\pm)}_{\sigma,\theta,{\vec p}}, 
\Omega^{(\pm)}_{\sigma',\theta',{\vec p}'})= \delta_{\sigma \sigma'}
\delta({\vec p}-{\vec p}')\delta(\theta-\theta')~,~~~
 (\Omega^{(\pm)}_{\sigma,\theta,{\vec p}}, 
\Omega^{(\mp)}_{\sigma',\theta',{\vec p}'})= 0~~.
\label{eq:E3.18}\end{eqnarray}   
These partial waves form a complete set (cf. Eq.~(\ref{eq:E3.32a})) and
therefore, can be used to decompose the fermion field,
\begin{eqnarray} 
\Psi (x)= \sum_\sigma \int d^2  {\vec p_t} d \theta 
[ \bbox{a}_{\sigma,\theta,{\vec p_t}}\Omega^{(+)}_{\sigma,\theta,{\vec
p_t}}(x)
+\bbox{b}^{\dag}_{\sigma,\theta,{\vec p_t}} 
\Omega^{(-)}_{\sigma,\theta,{\vec p_t}}(x)],\nonumber \\
\Psi^{\dag} (x)=\sum_\sigma\int d^2  {\vec p_t} d \theta 
[\bbox{a}^{\dag}_{\sigma,\theta,{\vec p_t}}
\overline{\Omega}^{(+)}_{\sigma,\theta, {\vec p_t}}(x)
+\bbox{b}_{\sigma,\theta,{\vec p_t}} 
\overline{\Omega}^{(-)}_{\sigma,\theta,{\vec p_t}}(x)]~.
\label{eq:E3.19}\end{eqnarray} 
The canonical quantization procedure, which identifies  the
coefficients $\bbox{a}$ and $\bbox{b}$ with the Fock operators, is standard,
and it leads to the anti-commutation relations,
\begin{eqnarray}
[\bbox{a}_{\sigma,\theta,{\vec p_t}}, 
\bbox{a}^{\dag}_{\sigma',\theta',{\vec p_t}'}]_+ =
[\bbox{b}_{\sigma,\theta,{\vec p_t}}, 
\bbox{b}^{\dag}_{\sigma',\theta',{\vec p_t}'}]_+ =
\delta_{\sigma \sigma'}
\delta({\vec p}-{\vec p}')\delta(\theta-\theta')~,
\label{eq:E3.19a}\end{eqnarray}
all other anticommutators being zero.
Non-trivial issues of the canonical quantization in wedge dynamics show up 
only in the gluon sector. They will be discussed in paper [III] of this cycle.

\section{Fermion correlators}
\label{sec:SN5}  

The field-theory calculations are based on various field correlators. A full
set of these correlators is employed by the Keldysh-Schwinger formalism 
\cite{Keld} which will be used below in the form given in
Refs.~\cite{QFK,QGD,tev}\footnote{The  indices of the contour ordering, as
well as the labels of linear combinations of variously ordered correlators, are
placed in
square brackets, e.g., $G_{[AB]}$, $G_{[ret]}= G_{[00]}-G_{[01]}$, etc.}. This
set consists of two Wightman  functions, where the field operators are taken in
fixed order,
\begin{eqnarray}
G_{[10]}(x_1,x_2)= -i\langle \Psi(x_1)\overline{\Psi}(x_2)\rangle ,~~~~
G_{[01]}(x_1,x_2)= i\langle \overline{\Psi}(x_2)\Psi(x_1)\rangle~,
\label{eq:E3.20}\end{eqnarray} 
and two differently ordered Green functions
\begin{eqnarray}
G_{[00]}(x_1,x_2)= -i\langle T[\Psi(x_1)\overline{\Psi}(x_2)]\rangle ,~~~~
G_{[11]}(x_1,x_2)= -i\langle T^\dag [\Psi(x_1)\overline{\Psi}(x_2)]\rangle~.
\label{eq:E3.21}\end{eqnarray} 
Here, $\langle\cdot\cdot\cdot\rangle$ denotes the average over an ensemble of 
the excited modes. The vacuum state (of each particular mode) is a part of
this
ensemble. In order to find the explicit form of these correlators, we shall
employ  the modes corresponding to the states with a given transverse
momentum $\vec{p_t}$ and an unusual rapidity  quantum number $\theta$.
Further on, it will be profitable to use the field correlators in the mixed
representation when they are Fourier-transformed only by their transverse
coordinates  $\vec{r_t}$, while the dependence on the proper time $\tau$
and the
rapidity coordinate $\eta$ is retained explicitly. Below, we derive the
corresponding expressions. The details of the derivation are important, since
they help to clarify physical issues related to the the localization of
quanta in the wedge dynamics, and are beneficial for the future analysis of
collinear singularities in paper [IV]. As for the ``vacuum part'' of the
correlators,
we obtain the more or less known expressions and put them into the form
which is convenient for future calculations.

For the practical calculations, we shall need not the functions $G_{[AB]}$
of Eqs. (\ref{eq:E3.20}) and (\ref{eq:E3.21}), but their linear combinations,
the fermion anti-commutator $G_{[0]}$ and the density of states $G_{[1]}$,
\begin{eqnarray}
G_{[0]}(x_1,x_2)= G_{[10]}(x_1,x_2)- G_{[01]}(x_1,x_2),~~~~
G_{[1]}(x_1,x_2)= G_{[10]}(x_1,x_2)+ G_{[01]}(x_1,x_2)~,
\label{eq:E3.20a}\end{eqnarray} 
and the retarded and advanced Green functions,
\begin{eqnarray}
G_{[ret]}(x_1,x_2)= G_{[00]}(x_1,x_2)- G_{[01]}(x_1,x_2) =
\theta (\tau_1-\tau_2)~G_{[0]}(x_1,x_2)~, \nonumber\\
G_{[adv]}(x_1,x_2)= G_{[00]}(x_1,x_2)- G_{[10]}(x_1,x_2)=
-\theta (\tau_2-\tau_1)~G_{[0]}(x_1,x_2)~.
\label{eq:E3.21a}\end{eqnarray} 
Nevertheless, we have to start with the computation of the simplest
correlators, the Wightman functions. Using Eqs.~(\ref{eq:E3.20}) and
(\ref{eq:E3.19}), we obtain
\begin{eqnarray}
G_{[10]}(x_1,x_2;p_t)=-i\int {d\theta \over 8\pi}
\big[ \gamma^+p_te^{-\theta}~e^{(\eta_1+\eta_2)/2}+
\gamma^-p_te^{+\theta}~e^{-(\eta_1+\eta_2)/2}
+p_r\gamma^r\gamma^0(\gamma^+e^{-{\eta_1-\eta_2\over 2}}
+\gamma^-e^{+{\eta_1-\eta_2\over 2}})\big]\nonumber\\
\times \bigg[[1-n^{+}(\theta,p_t)]
e^{-ip_t[\tau_1\cosh(\theta-\eta_1)- \tau_2\cosh(\theta-\eta_2)]}
+n^{-}(\theta,p_t)
e^{+ip_t[\tau_1\cosh(\theta-\eta_1)- \tau_2\cosh(\theta-\eta_2)]}\bigg]~.
\label{eq:E3.22}
\end{eqnarray}
It is useful to keep in mind a simple connection between this
expression and the  standard one.  Since, $\gamma^\pm=\gamma^0\pm\gamma^3$,
and $p_t e^{\pm\theta}=p^\pm\equiv p^0\pm p^3$, the first line in this
formula can be rewritten as $~\Lambda(-\eta_1)\not\! p ~\Lambda(\eta_2)$,
where 
\begin{eqnarray} 
\Lambda (\eta)=\cosh(\eta/2) + \gamma^0\gamma^3\sinh(\eta/2)=
{\rm diag}[e^{\eta/2},e^{-\eta/2},e^{-\eta/2},e^{\eta/2}]~, 
\label{eq:E3.23}\end{eqnarray}
is the matrix of Lorentz rotation with the boost $\eta$.  Furthermore, the
quantum number $\theta$ can be formally changed into $p_z$. Incorporating the
mass-shell delta-function $\delta(p^2)$ and returning to Cartesian
coordinates, we obtain,
\begin{eqnarray}  
G_{[10]}(x_1,x_2)
=\Lambda(-\eta_1)   \int {d^4 p \over (2\pi)^4} e^{-ip(x-x')} 
[ -2\pi i\delta (p^2)\not\! p]
\{(\theta(p^0)[1-n^{+}(p)] - (\theta(-p^0)n^{-}(p)\}\Lambda(\eta_2)~.   
\label{eq:E3.24}\end{eqnarray}
The expression between the two spin-rotating matrices $\Lambda$ is what is
commonly known for this type correlator in  flat Minkowsky space, and it
explicitly depends on the difference, $~x-x'~$,  of Cartesian coordinates.
The matrices $\Lambda(\eta)$ corrupt this invariance, because the curvature
of the hypersurface  of constant $\tau$ causes the effect known as Thomas
precession of the fermion spin that can be seen by an observer that changes
his rapidity coordinate and thus is subjected to an acceleration in
$z$-direction. From  the representation (\ref{eq:E3.24}), it is still
difficult to see that the correlators  (\ref{eq:E3.20}) depend only on the
difference $\eta=\eta_1-\eta_2$ (provided the distributions
$n^{\pm}(\theta,p_t)$ are boost invariant). This fact becomes clear after
we change the variable, $\theta = \theta'+(\eta_1+ \eta_2)/2$. Then
\begin{eqnarray} 
G_{[10]}(x_1,x_2;p_t)= - i\int{d\theta' \over 8\pi} 
\bigg[ \gamma^+~p_te^{-\theta'}+
\gamma^-~p_te^{+\theta'} +
p_r\gamma^r\gamma^0(\gamma^+e^{-{\eta\over 2}}
+\gamma^-e^{+{\eta\over 2}})\bigg]\nonumber\\ 
\times \bigg[[1- n^{+}\big({\eta_1+\eta_2\over 2}+ \theta',p_t\big)]
e^{-ip_t[\tau_1\cosh(\theta-\eta/2)- \tau_2\cosh(\theta+\eta/2)]}\nonumber\\ 
+ n^{-}\big({\eta_1+\eta_2\over 2}+ \theta',p_t\big)
e^{+ip_t[\tau_1\cosh(\theta-\eta/2)- \tau_2\cosh(\theta+\eta/2)]}\bigg]~. 
\label{eq:E3.25} \end{eqnarray} 
An amazing property of this formula is that the rapidity argument of the
distributions $n^{\pm}(\theta,p_t)$ is shifted by $(\eta_1+\eta_2)/2$ towards
the geometrical center of the correlator.  Now, the spin rotation in the 
$(tz)$-plane is virtually eliminated in such a way that both the spin 
direction and
the occupation numbers acquired a reference point exactly in the middle
between the endpoints $\eta_1$ and $\eta_2$. Now, things look  exactly
as if we had
performed the Wigner transform of the correlator. In actual fact, we did
not. If the distributions $n^{\pm}$ are boost-invariant along some finite
rapidity
interval, then the fermion  correlator (\ref{eq:E3.25}) will have the same
property.

The Wightman function (\ref{eq:E3.25}) has two different parts. One part is
connected with the vacuum density of states. The second ``material'' part is
connected with the  occupation numbers. The first one is always boost
invariant. Furthermore, we may expect that it depends (apart from the
spin-rotation effects) only on the invariant interval $\tau_{12}^2=
(t_1-t_2)^2-(z_1-z_2)^2$. The invariance of the material part is limited,
e.g., by the full width $2Y$ of the rapidity plateau and we have to be
careful in the course of further its transformation.   In order to extract
the dependence on $\tau_{12}$, we must make a second change of variable, 
$\theta' = \theta''+\psi$, where $\psi(\tau_1,\tau_2,\eta)$ depends on the
sign of the interval $\tau_{12}$ between the points $(\tau_1,\eta_1)$ and
$(\tau_2,\eta_2)$. Let the interval $\tau_{12}$ be time-like. Then 
\begin{eqnarray}
\tau_{12}^2=\tau_{1}^2+\tau_{2}^2 - 2\tau_1\tau_2\cosh\eta > 0,~~~~
\tanh\psi(\eta)=
{\tau_1+\tau_2\over\tau_1-\tau_2}\tanh{\eta\over 2},\nonumber\\ 
|\eta|<\eta_0 =\ln{\tau_1\over\tau_2}, ~~~
\tanh\psi(\pm\eta_0)=\pm1,~~~\psi(\pm\eta_0)=\pm \infty~. 
\label{eq:E3.26} 
\end{eqnarray} 
Then, Eq.~(\ref{eq:E3.25}) becomes 
\begin{eqnarray} 
G_{[10]}(x_1,x_2;p_t)= - i\int{d\theta'' \over 8\pi} 
\bigg[ \gamma^+~e^{-\psi}p_te^{-\theta''}+
\gamma^-~e^{\psi}p_te^{+\theta''} +
p_r\gamma^r\gamma^0\bigg(\gamma^+e^{-{\eta\over 2}}
+\gamma^-e^{+{\eta\over 2}}\bigg)\bigg]\nonumber\\ 
\times \bigg[[1- n^{+}\big({\eta_1+\eta_2\over 2}+\psi + \theta'',p_t\big)]
e^{-ip_t\tau_{12}\cosh\theta''}
+ n^{-}\big({\eta_1+\eta_2\over 2}+\psi + \theta'',p_t\big)
e^{+ip_t\tau_{12}\cosh\theta''}\bigg]~, 
\label{eq:E3.27} \end{eqnarray} 
and we see that the rapidity distributions of particles are shifted by 
$\psi(\eta)$ towards the direction between the points $(\tau_1,\eta_1)$ and
$(\tau_2,\eta_2)$.  According to (\ref{eq:E3.26}), the rapidity $\psi$ may be
infinite when this direction is light-like ($\tau_{12}^2=0$). Then this
shifted argument appears to be beyond the physical rapidity limits $\pm Y$
of the background distribution $n^{\pm}(\theta,p_t)$. This is extremely
important,  since this light-like direction is dangerous; it is solely
responsible for  the collinear singularities in various amplitudes.  One may
think that the cut-off $\pm Y$ will now appear as a parameter in the final
answer. This would be counter-intuitive, e.g., for many local quantities
related to the central rapidity region, like dynamical masses we are
intending to compute.  It will be shown later, that the theory is totally
protected from collinear singularities even in its vacuum part and no
explicit cut-off is necessary. 

For the case of a space-like  interval $\tau_{12}$, we introduce 
\begin{eqnarray}
\tilde{\tau}_{12}^2=-\tau_{12}^2=-\tau_{1}^2-\tau_{2}^2 +
2\tau_1\tau_2\cosh\eta >0~, \nonumber\\ 
\tanh\tilde{\psi}(\eta)=
{\tau_1-\tau_2\over\tau_1+\tau_2}\coth{\eta\over 2},~~~|\eta|>\eta_0~, 
\label{eq:E3.28} 
\end{eqnarray} 
and rewrite Eq.~(\ref{eq:E3.22}) as follows,
\begin{eqnarray} 
G_{[10]}(x_1,x_2;\vec{p_t})= - i\int{d\theta'' \over 4\pi} 
\bigg[ {1\over 2}\gamma^+~e^{-\tilde{\psi}}p_te^{-\theta''}+
{1\over 2}\gamma^-~e^{\tilde{\psi}}p_te^{\theta''} 
+p_r\gamma^r(\cosh{\eta\over 2}
-\gamma^0\gamma^3\sinh{\eta\over 2})\bigg]\nonumber\\ 
\times \bigg[ [1- 
n^{+}\big({\eta_1+\eta_2\over 2}+\tilde{\psi} + \theta'',p_t\big)]
e^{-ip_t\tilde{\tau}_{12}{\rm sign}\eta\sinh\theta''}+
n^{-}\big({\eta_1+\eta_2\over 2}+\tilde{\psi} + \theta'',p_t\big)
e^{+ip_t\tilde{\tau}_{12}{\rm sign}\eta\sinh\theta''}\bigg]~. 
\label{eq:E3.29} \end{eqnarray} 
Now it is easy to see that we are protected from the null-plane singularities 
in the material sector of the theory on both sides of the light-like plane.
Similar calculations can be done for the second Wightman function $G_{[10]}$
which differs from $G_{[10]}$ by the obvious replacements, $1-n^+\to -n^+$, and
$-n^-\to 1-n^-$. The results can be summarized as follows,
\begin{eqnarray}
 G_{[10]}(\tau_1,\tau_2,\eta;\theta'', p_t) =[1-n^+(\theta,p_t)]
 G^{(0)}_{[10]}(\tau_1,\tau_2,\eta;\theta'', p_t) -
 n^-(\theta,p_t) 
 G^{(0)}_{[01]}(\tau_1,\tau_2,\eta;\theta'',p_t)~~,\nonumber\\
 G_{[01]}(\tau_1,\tau_2,\eta;\theta'', p_t) = -n^+(\theta,p_t)
 G^{(0)}_{[10]}(\tau_1,\tau_2,\eta;\theta'', p_t) + [1-n^-(\theta,p_t)]
 G^{(0)}_{[01]}(\tau_1,\tau_2,\eta;\theta'', p_t)~, 
\label{eq:E3.30}
\end{eqnarray}
where according to Eqs.~(\ref{eq:E3.25}), (\ref{eq:E3.27}) and
(\ref{eq:E3.29}),
$\theta=(\eta_1+\eta_2)/ 2+\psi + \theta''$.
Here, $G^{(0)}_{[\alpha]}$ is the vacuum counterpart of  $G_{[\alpha]}$,
and $~G^{(0)}_{[01]}(x_1,x_2;\theta,\vec{p_t})= 
[G^{(0)}_{[10]}(x_1,x_2;\theta,-\vec{p_t})]^\ast$.
Using Eqs.~(\ref{eq:E3.30}), we may easily obtain the field correlators
defined by Eqs.~(\ref{eq:E3.20a}).
One of them is the causal anti-commutator,
\begin{eqnarray}
G_{[0]}(x_1,x_2;\vec{p_t})\equiv
G_{[10]}(x_1,x_2;\vec{p_t})- G_{[10]}(x_1,x_2;\vec{p_t})=
G^{(0)}_{[10]}(x_1,x_2;\vec{p_t})- G^{(0)}_{[10]}(x_1,x_2;\vec{p_t})~,
\label{eq:E3.31}\end{eqnarray} 
which does not include occupation numbers, while the density of states,
\begin{eqnarray}
G_{[1]}(x_1,x_2;\vec{p_t})\equiv 
G_{[10]}(x_1,x_2;\vec{p_t})+ G_{[10]}(x_1,x_2;\vec{p_t})
= [1-2n_f(\theta,p_t)]G^{(0)}_{[1]}(x_1,x_2;\vec{p_t})~,
\label{eq:E3.32}\end{eqnarray} 
carries all the information about the phase-space population. In the last
equation, we have put $n^-(\theta,p_t)=n^+(\theta,p_t)=n_f(\theta,p_t)$.
It is straightforward to check that 
\begin{eqnarray}
G_{[0]}(\tau,\eta_1,\vec{r}_{t1};\tau,\eta_2,\vec{r}_{t2} )
= -i\langle 0|~[\Psi(\tau,\eta_1,\vec{r}_{t1}),
\overline{\Psi}(\tau,\eta_2,\vec{r}_{t2} )]_+ ~|0\rangle=
-i{\gamma^0\over\tau}\delta(\eta_1-\eta_2)\delta(\vec{r}_{t1}-\vec{r}_{t2})~.
\label{eq:E3.32a}\end{eqnarray} 
This property of the equal-proper-time commutator is the canonical commutation
relation which is translated into commutation relations for the Fock
operators.
It also verifies that the system of wave functions we employ forms a complete
set.

In any calculations connected with the local quantities in heavy ion
collisions, we would like to rely on the rapidity plateau in all
distributions and to avoid its width as a parameter in the final answers.
If this is possible (which appears to be the case), then we may consider
the occupation numbers as the functions of $p_t$ only, and accomplish the
integration over the rapidity $\theta''$. This integration gives the vacuum
correlators $G_{[\alpha]}(\tau_1,\tau_2,\eta; \vec{p_t})$ in  closed
form.\footnote{Some of the integrals over $\theta''$ are defined as
distributions by means of analytic continuation.} The integrations are
straightforward and result in the following representation of the fermion
correlators,
\begin{eqnarray}
G_{[\alpha]}(\tau_1,\tau_2;\eta,\vec{p_t})=
\gamma^+ ~p_t g^{L(+)}_{[\alpha]} +\gamma^- ~p_t g^{L(-)}_{[\alpha]} +
p_r\gamma^r\gamma^0\gamma^+ ~g^{T(+)}_{[\alpha]} + 
p_r\gamma^r\gamma^0\gamma^- ~g^{T(-)}_{[\alpha]}~.
\label{eq:E3.33}
\end{eqnarray}
The products of three gamma-matrices in this expression indicates that the
fermion correlators acquire an axial component ($\sim \gamma^r\gamma^5$),
which is consistent with the absence of complete Lorentz and rotational 
symmetry in our problem.  In order to obtain the compact expressions for the
invariants $g_{[\alpha]}$, one must note that in all domains, we can replace 
\begin{eqnarray}
\bigg( ~e^{\mp\psi},~ \mp{\rm sign}\eta ~ e^{\mp \tilde{\psi}} ~\bigg)
~\to ~ {\tau_1 e^{\mp\eta/2}-\tau_2e^{\pm\eta/2}\over \sqrt{|\tau_{12}^2|}}~.
\label{eq:E3.34} 
\end{eqnarray} 
These transformations lead to the  final  expressions for the invariants of
the fermion correlators that we shall use in our calculations. For the
invariants of the causal anti-commutator $G_{[0]}$, we have
\begin{eqnarray}
 g^{L(\pm)}_{[0]}=
 i~{\tau_1 e^{\mp\eta/2}-\tau_2 e^{\pm\eta/2}\over 4\sqrt{|\tau_{12}^2|}}
 \theta(\tau_{12}^2)~J_1(p_t\tau_{12})~,~~~~
 g^{T(\pm)}_{[0]}= -~{ e^{\mp\eta/2}\over 4}
 \theta(\tau_{12}^2)~J_0(p_t\tau_{12})~.
\label{eq:E3.35} 
\end{eqnarray} 
They are causal in the sense,  that they are completely confined to the
interior of the future light wedge. Depending on the context, the invariants of
the density $G_{[1]}$ will be used in two different representations, 
\begin{eqnarray}
  g^{L(\pm)}_{[1]}=  - \int{d\theta' \over 4\pi} 
\big[1- 2 n_f\big({\eta_1+\eta_2\over 2}+ \theta',p_t\big)\big]  
~e^{\mp\theta'}
~\sin\big( p_t[\tau_1\cosh(\theta-\eta/2)- 
   \tau_2\cosh(\theta+\eta/2)]\big)\nonumber\\
={\tau_1 e^{\mp\eta/2}-\tau_2e^{\pm\eta/2}\over 4\sqrt{|\tau_{12}^2|}}
  \bigg[\theta(\tau_{12}^2)~Y_1(p_t\tau_{12})
  +{2\over\pi}\theta(-\tau_{12}^2)
  K_1(p_t\tilde{\tau_{12}})\bigg]
~\big[ 1-2n_f(p_t) \big]~,
\label{eq:E3.36}
\end{eqnarray}
\begin{eqnarray}
g^{T(\pm)}_{[1]}= -i e^{\mp\eta/2} \int{d\theta' \over 4\pi}
\big[1- 2 n_f\big({\eta_1+\eta_2\over 2}+ \theta',p_t\big)\big]
~\cos\big( p_t[\tau_1\cosh(\theta-\eta/2)- 
   \tau_2\cosh(\theta+\eta/2)]\big)\nonumber\\
= i{ e^{\mp\eta/2}\over 4}
  \bigg[\theta(\tau_{12}^2)~Y_0(p_t\tau_{12})
  -{2\over\pi}\theta(-\tau_{12}^2)
  K_0(p_t\tilde{\tau_{12}})\bigg]~\big[ 1-2n_f(p_t) \big]~.
\label{eq:E3.37}
\end{eqnarray}
The first of these representations will be expedient when the quark from the
distribution $n_f(p_t)$ in the self-energy loop is interacting with the
radiation component of the gluon field which imposes the physical limits on
the rapidity $\psi(\eta)$ in the phase, $\Phi=\tau_{12}(\eta) p_t \cosh
(\theta'-\psi(\eta))$, in the integrands of Eqs.~(\ref{eq:E3.36}) and
(\ref{eq:E3.37}). Then, the integration $d\theta'$ will have finite limits
defined by the light cone and the localization of states with the large
$p_t$. When the quark interacts with the longitudinal (static) component of
the gluon field, no limitations of this kind appear and we are able to use
the second analytic representation.

\vspace{1cm}

\noindent {\bf ACKNOWLEDGMENTS}

The author is grateful to Berndt Muller, Edward Shuryak and Eugene Surdutovich
for helpful discussions at various stages in the development of this work,
and appreciate the help of Scott Payson who critically read the
manuscript.

This work was partially supported by the U.S. Department of Energy under
Contract  No. DE--FG02--94ER40831.


\bigskip


\begin{references}
\bibitem{QFK} A. Makhlin, Phys.Rev. {\bf C 51} (1995) 34
\bibitem{QGD} A. Makhlin, Phys.Rev. {\bf C 52} (1995) 995.
\bibitem{tev}  A. Makhlin and E. Surdutovich, Phys.Rev. {\bf C 58} (1998) 389
           (quoted as paper [I]).
\bibitem{Gribov} V.N. Gribov, {\em Space-time description of hadron
              interactions at high energies}, in Proceedings of the 8-th
              Leningrad Nuclear Physics Winter School, February 16-27, 1973.
\bibitem{BPQCD}  Yu.L. Dokshitzer et al., {\em Basics of perturbative QCD},
                 Editions Frontieres, 1991.    
\bibitem{Shuryak} E.V. Shuryak, Sov. Phys. JETP {\bf 47} (1978) 212.
\bibitem{geg}   A. Makhlin, {\em Scenario for Ultrarelativistic Nuclear 
                Collisions: III.~ Gluons in the expanding geometry.}   
		hep-ph/0007301 (quoted as paper [III]) 
\bibitem{fse}   A. Makhlin and E. Surdutovich, 
                {\em Scenario for Ultrarelativistic Nuclear Collisions:
                IV.~Effective quark mass at the early stage.}  
		hep-ph/0007302(quoted as paper [IV]) 
\bibitem{Sakharov}  A.D. Sakharov, JETP {\bf 18}, 631  (1948). 
\bibitem{Landau} L.D. Landau, E.M. Lifshits, Quantum mechanics, Sec.1, 
               Oxford ; New York : Pergamon Press, 1977
\bibitem{Dirac} P.A.M. Dirac, Rev.Mod. Phys, {\bf 21}, 392  (1949). 
\bibitem{Fock}   V.A.Fock, Z. f. Phys. {\bf 57},  261 (1929).
\bibitem{Witten}  M.B. Green, J.H. Schwarz, E. Witten, Superstring
                theory, Cambridge University press, 1987.
\bibitem{Keld}  L.V. Keldysh, Sov. Phys. JETP {\bf 20} (1964) 1018; 
               E.M. Lifshits, L.P. Pitaevsky, Physical kinetics, 
             Pergamon Press, Oxford, 1981.              
\end{references}
\end{document}